\documentclass[10pt,twocolumn,letterpaper]{article}

\usepackage{iccv}
\usepackage{times}
\usepackage{epsfig}
\usepackage{graphicx}
\usepackage{amsmath}
\usepackage{amssymb}
\usepackage{multirow}

\usepackage[accsupp]{axessibility}  

\usepackage[breaklinks=true,bookmarks=false]{hyperref}

\iccvfinalcopy 

\def\iccvPaperID{5} 

\ificcvfinal\fi

\begin{document}

\title{A 3D CNN Network with BERT For Automatic COVID-19 Diagnosis From CT-Scan Images}

\author{Weijun Tan*, Jingfeng Liu\\
Shenzhen Deepcam Information Technologies, China\\
{\tt\small \{weijun.tan,jingfeng.liu\}@deepcam.com} \\
*LinkSprite Technologies, USA\\ \tt\small  weijun.tan@linksprite.com
}

\maketitle
\ificcvfinal\thispagestyle{empty}\fi

\begin{abstract}
   We present an automatic COVID1-19 diagnosis framework from lung CT-scan slice images. In this framework, the slice images of a CT-scan volume are first preprocessed using segmentation techniques to filter out images of closed lung, and to remove the useless background. Then a resampling method is used to select a set of fixed number of slice images for training and validation. A 3D CNN network with BERT is used to classify this set of selected slice images. In this network, an embedding feature is also extracted. In cases where there are more than one set of slice images in a volume, the features of all sets are extracted and pooled into a feature vector for the whole CT-scan volume. A simple multiple-layer perceptron (MLP) network is used to further classify the aggregated feature vector. The models are trained and evaluated on the provided training and validation datasets. On the validation dataset, the precision is 0.9278 and the F1 score is 0.9261. On the test dataset, the F1 score is 0.8822. The code is available at \url{https://github.com/deepcam-cn/3D-CNN-BERT-COVID19}. 
\end{abstract}

\section{Introduction}

The novel COVID-19 coronavirus breaking out in late 2019 has been one of the worst disasters in the human history \footnote{https://covid19.who.int/}. Therefore, it is very critical to stop the spreading of the virus. So first a person has to be confirmed to have COVID-19 before safety measures and treatments can be taken accordingly. One example is \cite{COVID-trace}, where thermal imaging is used to detect fever patients and face recognition is used to report and trace patients and their close contacts. 

Among the techniques to diagnosis COVID-19, X-ray and CT-scan images are studied extensively. In this paper, we present an automatic diagnosis framework from chest CT-scan slice images. The goal is to classify COVID-19, and non-COVID-19 from a volume of CT-scan slice images of a patient. We use the dataset provided in the ICCV2021 MIA-COV19D challenge \footnote{https://mlearn.lincoln.ac.uk/mia-cov19d/}  \cite{kollias2021mia}. Since there is no slice annotation in this dataset, 2D CNN classification network on single slice image is not considered. Instead, 3D CNN network is explored.  

In this paper, we first discuss the preprocessing of CT-scan slice images, which is very critical for good classification performance. The goal of preprocessing is to prepare good-quality slice images per volume for training and validation in the CNN classification network. In this work, we propose segmentation techniques using both traditional morphological transforms and a deep learning method using UNet \cite{unet}. Since the 3D CNN network requires fixed number of images as input, we adopt and revise a resampling method used in \cite{He2021CovidNet3D}. In validation and testing time, we propose a method to use all available good slice images to make the final classification prediction.   

A 3D network is widely used in many tasks including video understanding (recognition, detection, captioning). In videos, frame images at different time form a 3D series of images. A 3D CNN network is good at aggregating the temporal information. 3D network is also used in COVID-19 diagnosis, where the slice images at different spacing form a 3D series of images. The correlation between slice images is just analogous to the temporal information in videos. We are the first to use a 3D CNN network with BERT (CNN-BERT) for video action recognition \cite{bert} in classification of COVID-19 from CT-scan images.  

Most 3D CNN networks use fixed number of images as input. Others can use all available images by a global pooling method.  In our study, to avoid the out-of-memory problem caused by using all slice images, we choose to use fixed number of images as input to the 3D CNN-BERT network. However, since the available images may be a lot more than this fixed number, if only one set of them is used, some useful information may be missed. Motivated by this insight, we extract the embedding feature vector of all available sets of images for every CT-scan volume, then aggregate one global feature vector out of them. This feature vector is sent to a simple MLP classifier for extra classification. This way, both the advanced 3D CNN network with BERT and the information in all available image are used. Evaluation results show that this pooling method and MLP can improve the accuracy on the validation dataset by 1.6\%. To our knowledge, we are the first to explore using embedding features of all available sets of slice images in a 3D CNN network for classification of COVID-19.          

On the provided validation dataset, we achieve an accuracy 0.9278, and an F1 score 0.9261. On the test dataset, we achieve an F1 score 0.8822.    

\section{Related Work}

Deep learning has been used in a lot of medical imaging analysis and diagnosis, e.g. in  \cite{kollias2018deep} \cite{kollias2020deep} and \cite{kollias2020transparent}. 

Since the outbreak of the COVID-19 pandemic, a lot of researches have been done to diagnose it using deep learning approaches, mostly CNNs on CT scan images or X-ray images. For a complete review, please refer to \cite{Review1} and \cite{Review2}. 

Among the classification methods, some use 2D network on slice image individually and make prediction for every image. This is called 2D network. To make a decision for a patient, some kinds voting method is typically used \cite{SPGC1}, \cite{COVID-CTSet}, \cite{COVID-FACT}. Others use 2D network on slice image, and generate embedding feature vector for every image, then all feature vectors are pooled to a single global feature vector, and a few fully-connection (FC) layers are used for classification. This is called 2D+1D network \cite{li2020artificial}, \cite{CTCAPS}, \cite{SPGC2}. The third method is a pure 3D CNN network, where slice annotation is not needed, and a set of or all the available slice images are used as input, and the 3D network process all these input images all at once in a 3D channel space \cite{Tongji}, \cite{SPGC3}, \cite{He2021CovidNet3D}. 

In the 2D CNN method, some use the lung mask segmentation, but most of them directly use the raw slice image.  The COVID-MaskNet \cite{COVID-CT-Mask-Net} uses a segmentation network to localize the disease lesion, then use a FasterCNN-based approach to do the classification on the detected lesion regions. The COVID-Net Initiative \cite{Gunraj2020}, \cite{Gunraj2021} have done extensive studies of COVID classification on both CT scan images and X-ray images. They also collect and publish the largest CT image dataset - so called COVIDx CT-2 dataset. In \cite{COVID-CTSet}, Resnet50 with FPN is used. In \cite{SPGC1}, a combination of infection/non-infection classifier, and a COVID-19/CAP/normal classifier is used. 

In the 2D+1D method, in \cite{li2020artificial}, a pretrained 2D Resnet network is used to extract a feature vector out of every image, then all the features are pooled using max-pooling. This feature is sent to a few FC layers to make classification prediction.  In \cite{CTCAPS}, a Capsule network is used to extract feature vector for every image, then these feature vectors are pooled using max-pooling into a global feature vector and a decision is made for the volume. In \cite{SPGC2}, a feature vector is extracted for every image, then multiple pooling methods are ensembled to generate a global feature vector before a classification is made. In \cite{kollias2021mia}, an RNN is used to aggregate 2D features.   

In the 3D CNN method, in \cite{Tongji}, a 3D CNN network is used, however, with both the slice image and a segmented lung mask as input. They discard a fixed percentage of slice images at the beginning and end of a CT-scan volume. In \cite{SPGC3}, the authors first segment the lung mask from a slice image using traditional morphological transforms, then use this mask to select good slice images and generate lung-only images (no bone or tissue) slice images. To make the number of images a fixed number, they use 3D cubic interpolation to regenerate slice images. In \cite{He2021CovidNet3D}, 3D CNN network using a fixed number of slice images as input is used. Instead of using a fixed 3D CNN architecture, an autoML method is used to search for best 3D CNN architecture in the network space with MobileNetV2 \cite{MobileNetV2} block. 

\section{Data Preprocessing and Preparation}

The first goal of preprocessing is to select good slice images for training and validation. The second goal is to segment the lung mask of a slice image, so a masked image, where background, bones, vessels, tissues are all blacked out. This has been shown useful in \cite{CTCAPS}, \cite{COVID-FACT}, \cite{SPGC3}.     

\subsection{Morphological Lung Segmentation}

In this work, we use two segmentation techniques to segment the lung region out of the slice image. The First technique is based on traditional morphological transforms. It works well to get an bounding box of the lung (including bones and tissues), and a coarse mask image of the two lungs. The bounding box can be used to remove background of the slice image, and the mask can be used to tell if the lungs are closed. This bounding box and the mask image are used to select slice images of a patient for training and validation.  

The segmentation involves Gaussian blurring, binarization, erosion and dilation, contour finding, seed filling, clearing border, labelling, filling holes etc.  Shown in Figure 1 (a)-(c) is an example of this segmentation. In this example, the raw image has infection lesions. In the lung mask, many important parts are filtered out by this segmentation.     

\begin{figure}[t]
    \centering
    \includegraphics[scale=0.23]{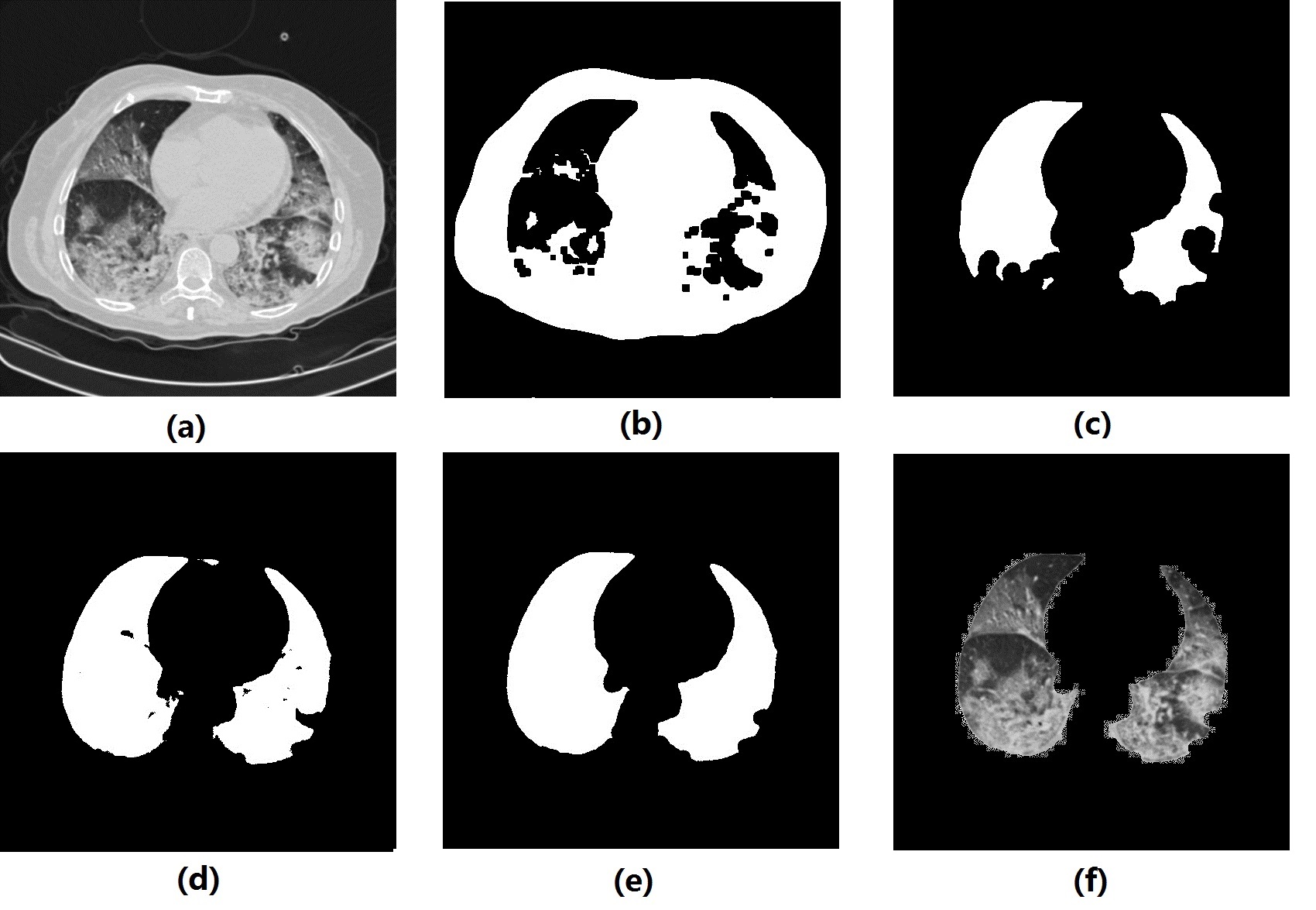}
    \caption{Example of lung segmentation: (a) raw image, (b) after binarization, (c) morphological segmented mask, (d) original UNet segmented mask, (e) refined UNet segmented mask, (f) masked lung image}.
    \label{fig1}
\end{figure}

\subsection{UNet Segmentation}

The morphological lung segmentation may miss many important details of the true lung mask, particularly near the edges, and on images with infection lesions. So it does not work well for segmenting the lung images for COVID-19 classification. To overcome this problem, a UNet segmentation network \cite{unet} is trained. The datasets from Kaggle \footnote{https://www.kaggle.com/kmader/finding-lungs-in-ct-data} and CNBC \cite{CNCB} are used. Since the ICCV-MIA dataset \cite{kollias2021mia} does not have any lung mask annotations, we do not retrain the model. We simply use the trained UNet model to segment the slice images in the ICCV-MIA dataset. Furthermore, since there are holes and noisy boundaries in the segmented mask, we use further morphological transforms to smooth the edges and fill the holes. The mask image after this step is used to extract the lung image for training and validation. Shown in Figure 1 (d)-(e) is example of the original and refined UNet masks. Figure 1 (f) is the masked lung image.   

Generally speaking, the UNet segmentation works better than the morphological segmentation. However, it works poorly on the closed-lung images. That is why we still need the morphological segmentation for the selection of slices images. 

\subsection{Selection of Slice Images}

Previous works show that some slice images, particularly those of closed lungs, are useless in classification the CT-scan volume \cite{COVID-CTSet}, \cite{Tongji}, \cite{SPGC3}. Therefore, it is beneficial to filter out these closed-lung slice images. In \cite{Tongji}, a fixed percentage of slice images at the beginning and end of a volume is discarded. In \cite{SPGC3}, lung mask is segmented, and the images whose percentages of lung mask in the whole image is less than a threshold are filtered out.       
We use a method similar to \cite{SPGC3}. However,since some volumes have very small number of slice images, we make the threshold adaptive. In the first step, the threshold 0.7 is used, i.e., slice images whose percentage of lung mask is less than 0.7 of the largest percentage in the whole volume are filtered out. In the next step, if the remaining number of slice images is too few (e.g., 8), then we reduce the threshold by 0.1 until 8 images remain or the threshold reaches 0. 

\subsection{Slice Images Resampling}

As we explain before, we use fixed number (32) of slice images as the input to the 3D CNN-BERT network. However, since the number of available slice images is varying, we need to use resampling strategy to generate the input slice images. There are two cases down-sampling and up-sampling. 
We use the resampling idea similar to \cite{He2021CovidNet3D}. On the training dataset, random sampling is used, while on the validation and test datasets, a symmetrical and uniform resampling is used. 

In the training and validation time, only one set of images is selected in every epoch. However, in the testing time, we use a different symmetrical resampling method. Instead of selecting one set of data, we select multiple sets of data if there are enough available images. 

\subsection{Input Image Composition}

The 3D CNN-BERT network \cite{bert} requires to have three-channel input image, which is typically RGB. In CT-scan, all slice images are in gray-scale.  So we need to compose three-channel input images.  

The first choice is the raw gray-scale slice image, which is denoted as an R channel, as shown in Figure 1 (a). In \cite{Tongji}, the authors propose to use the segmented mask as part of the input. We use it as well as one choice in our design, which is denoted as an M channel, as shown in Figure 1 (e). In many works, the lung image masked by the mask, i.e., the pixels inside the lung mask keep their gray-scale values, while all other pixels have value 0, are used as input to the classification network. We use this lung image as the third choice, which is denoted as the L channel, as shown in Figure 1 (f). So in comparison to the RGB image, we have RML image, where the channels R, M, L can be combined in many different ways. We will show different performance of these images.       

Since we have the bounding box for every image in a CT-scan volume, we take the largest bounding box of all images as the bounding box for the volume. This bounding box can be used to crop the RML image inside the bounding box. 

\section{Classification Networks}

In this paper, we explore two levels of 3D classification. In the first level, a 3D CNN-BERT network \cite{bert} is used. In the second level, feature vectors of all available set of slices images are pooled to a global feature vector for every CT-scan volume. This feature vector is sent to a simple MLP classifier for second level classification. 

\subsection{3D CNN-BERT Network}

We reuse the 3D CNN-BERT network in \cite{bert}, as shown in Figure 2. This architecture utilizes BERT-based temporal pooling for video action recognition. In this work, we use it to pool the correlation between CT-scan slice images. In this architecture, the selected 32 slice images from the resampling scheme are propagated through a 3D CNN architecture without applying temporal global average pooling at the end of the architecture. A learned positional encoding is added to the extracted features. In order to perform classification with BERT, additional classification embedding (xcls) is appended (represented as red box in Figure 2). The classification of the architecture is implemented with the corresponding classification vector ycls which is sent the FC layer, producing the classification output. 

As the authors point out, the use of the temporal attention mechanism for BERT is not only to learn the convenient subspace where the attention mechanism works efficiently, but also to learn the classification embedding which learns how to attend the temporal features of the 3D CNN architecture properly \cite{bert}. In \cite{bert} many 3D backbone networks are studied, including the R(2+1)D \cite{r2+1d}. On two main video action recognition benchmark datasets, this architecture achieves state-of-the-art performance. For more detail of this architecture, including the cost function, please refer to \cite{bert}.     
  
In this paper, we use the R(2+1)D backbone \cite{r2+1d} where the Resnet34 \cite{Resnet} is used. We use 32 slice images as input.  We make some modifications, including using input image size 224x224, while 112x112 is used in \cite{bert}, and outputting the embedding feature. We use very different data augmentations from \cite{bert}, which are proven effective to bring the best performance.  

\begin{figure}[t]
    \centering
    \includegraphics[scale=0.375]{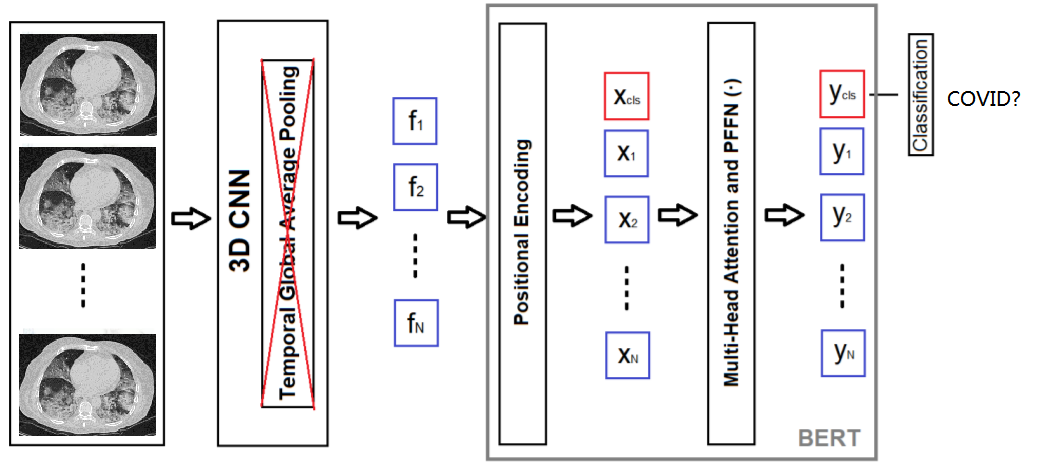}
    \caption{The modified 3D CNN-BERT Architecture from \cite{bert} }
    \label{fig2}
\end{figure}

\subsection{MLP Classification Network}

The 3D CNN-BERT network produces a classification result for a single set of input slice images. For CT-volumes where there are more than one set of slice images, we want to process all available slice images in order not to miss some useful information. Therefore, we propose to use a second level MLP classifier on the pooled embedding feature vectors out of the 3D CNN-BERT network.  

To aggregate a global feature vector from multiple local feature vectors, the max-pooling and the avg-pooling are two most popular methods. We consider both of them, and a concatenation of them. The first FC1 layer has 128 neurons, and the second FC2 layer has 32 neurons. We consider different activation functions, including ReLU and Sigmoid after each FC layer. In order to prevent overfitting, we use drop out of 0.5 after first two FC layers. This MLP network is depicted in Figure 3.  

\begin{figure}[t]
    \centering
    \includegraphics[scale=0.7]{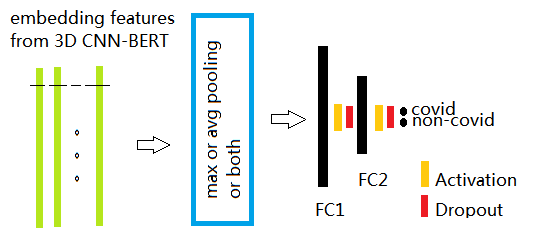}
    \caption{The MLP classification network}
    \label{fig3}
\end{figure}

\section{Experiment Results}

\subsection{Dataset}

In the ICCV-MIA dataset \cite{kollias2021mia}, there are 1552 training CT-scan  volumes, 374 validation volumes, and 3455 test volumes. In each volume, there are various number of slice images, ranging from 1 to more than a thousand. There are no slice annotations, so 2D CNN classification is not possible if not using extra datasets. 

Most of the sizes of the images are 512x512. However, there are quite a lot images whose sizes are not so. So we first do a quick check, if the size of an image is not 512x512, it is resized to 512x512 before any other preprocessing is applied.

For the Unet segmentation, we use the annotated dataset we find on Kaggle and from CNBC \cite{CNCB}. In the CNBC annotation, three types of annotations - lung field, Ground-glass opacity, and Consolidation. We merge all three types to the lung field.  

No other datasets are used in training or validation of the 3D CNN-BERT network or the MLP network.  

\subsection{Implementation details}

The 3D CNN-BERT is implemented with Pytorch framework. Input image size is set to 224x224. In order to do so, a few changes are made in the source code from \cite{bert}. We choose to use 32 slice images as input and use the R(2+1)D backbone. We use the Adam optimizer with a initial learning rate 1E-5. A reduced learning rate on plateau with a factor 0.1 and 5 epoch patience is used. The training runs at most 200 epochs with early stopping. The validation accuracy is used to select the best model.   

The MLP network is also implemented with Pytorch. The input feature vector size is 512 or 1024. We use the Adam optimizer with a initial learning rate 1E-4 or 1E-5. A reduced learning rate on plateau with a factor 0.1 and 5 epoch patience is used. The training runs at most 100 epochs with early stopping. The validation accuracy is used to select the best model.     

The UNet segmentation network is implemented with Keras-2.3 and Tensorflow-GPU-2.2. The input image size is 512x512. We use the Adam optimizer with a initial learning rate 1E-5. A reduced learning rate on plateau and early stopping are used. The validation loss is used to select the best model. 

\textbf{Data Augmentations}: To make the model generalize well, data augmentation is very important.  In this work, on the training dataset, we use random Affine transforms with rotation of 10 degree, scaling range of 0.8-1.2, translation range of 0.1, width shearing of 10 degree, brightness of 50\%, contrast of 30\% \cite{Tongji}. In addition, the multiple scale cropping (MSC) \cite{multiscalecrop} is used with the image enlarged by 25\%. Random horizontal flipping is also used. On the validation and test dataset, central cropping is used with the image enlarged by 25\%.  In the ablation study we will show some results using different data augmentation methods.      

\subsection{Ablation Studies}

\subsubsection{3D CNN-BERT Network}

In this section we compare the performance of different configurations on the validation dataset.  

\textbf{Input Image Composition and Bounding Box Cropping}: In first panel of Table 1 we show the validation accuracy of different image compositions without or with bounding box cropping. In these tests, the MSC and random horizontal flipping augmentations are used.  

\textbf{Data Augmentation}: We use the best configuration from the previous tests, and test the effects of data augmentation. The most important factor is to find out if the random Affine transforms help the accuracy. The results are shown in the second panel of Table 1.

From these studies, we find that the RML image with MSC augmentation without bounding box cropping gives the best accuracy performance. 

\textbf{3D CNN without BERT}: We use the best configuration from the previous tests, and test the effect of 3D CNN network without BERT. The result is listed in the third panel of Table 1. It is a surprise to see that without BERT, the 3D CNN performs very poorly. We also try different 3D network, like the resneXt3D \cite{resneXt3D}, the accuracy is around 0.71. This shows the power of the BERT on classifying COVID-19.          

\begin{table}
	\begin{center}
		\begin{tabular}{cccc}
		    \hline
			Input & Augmentation  & Crop? & Acc(\%)\\
			\hline
			RRR & MSC & No & 89.06\\
			LLL & MSC & No & 89.84\\
			RRL & MSC & No & 90.10\\
			RML & MSC & No & \textbf{91.18}\\
			RML & MSC & Yes & 87.43\\
            \hline
			RML & MSC+A & No & 90.91\\
			RML & A & No & 87.43\\
			\hline
			RML-NoBERT & A & No & 68.45\\
			\hline
		\end{tabular}
		\caption{3D CNN-BERT classification accuracy on validation dataset for different configurations. In augmentation, two important approaches Affine transforms (A) and multiple scale cropping (MSC) are studies.} 
		\label{T1}		
	\end{center}
\end{table}

\subsubsection{MLP Network}
We use the RML image with the Affine transform and multiple-scale cropping of Table 1 in this test. Embedding features of all training and validation volumes are extracted and save as Pickle files.  

We test the max-pooling, avg-pooling, and concatenation of both. For the activation function, we test ReLU, Sigmoid, and Tanh functions. The results are listed in Table 2. We see that the concatenation of the max-pooling and avg-pooling gives better results than using only one of them. Out of the three activation functions, the Sigmoid gives best performance but takes longer time to converge or needs to use larger initial learning rate 1E-4.  

Furthermore, we find that adding a new set of selected images at the center of a CT-scan volume to previous symmetrical and uniform resampling can help improve the MLP performance, so we include it in the final benchmarking tests. 

\begin{table}
	\begin{center}
		\begin{tabular}{cccc}
		    \hline
			Pooling & Activation & Accuracy(\%)\\
			\hline
			 -  & -  & 91.18 (baseline)\\
			\hline
			Max & ReLU & 91.44\\
			Avg & ReLU & 90.91\\
			Both & ReLU & 91.18\\
			Max & Sigmoid & 91.71\\
			Avg & Sigmoid & 91.18\\
			Both & Sigmoid & \textbf{91.98}\\
			Both & Tanh & 91.72\\
			\hline
		\end{tabular}
		\caption{MLP classification accuracy on validation dataset.} 
		\label{T2}		
	\end{center}
\end{table}

\subsection{Benchmark Results}

After the ablation study, we choose a few top performers to bench mark our algorithm. On the ICCV-MIA \cite{kollias2021mia} validation dataset, the benchmark results are listed in Table 3. In this table, both the accuracy and the F1 score are presented.   

On the ICCV-MIA test dataset \cite{kollias2021mia}, the best F1 scores from the 3D CNN-BERT and the MLP classification networks are 88.22\% and 87.4\% respectively.  

\begin{table}
	\begin{center}
		\begin{tabular}{ccc}
		    \hline
			Dataset & 3D-CNN-BERT & MLP\\
			\hline
			Validation & (91.18,91.00) & \textbf{(92.78,92.61)} \\
			Validation & (90.91,91.65) & (92.25,92.05) \\
			Test & (-,88.22) & (-,87.4) \\
			\hline
		\end{tabular}
		\caption{Classification results (accuracy \%,F1-score \%) on the validation and test datasets.} 
		\label{T3}		
	\end{center}
\end{table}

\section{Conclusions}

In this paper we present a 3D CNN-BERT network with an extra MLP network for COVID-19 classification. The MLP can improved the accuracy by 1.6\%. On the validation dataset, our best F1 score is 92.61\%. And on the test dataset, our best F1 score is 88.22\% from the 3D CNN-BERT, while the MLP does not bring extra improvement.     

{\small
\bibliographystyle{ieee_fullname}
\bibliography{iccvcovid}
}

\end{document}